\documentclass[10pt]{amsart}
\usepackage{amsthm, hyperref,amscd, graphicx, xy, tikz, color}
\usepackage{amsthm}
\numberwithin{equation}{section}

\theoremstyle{definition}

\title[Benjamin--Feir instability of Wave Packets]{Benjamin--Feir instability of Wave Packets at Interface of Liquid Half-Space and Layer}
\author[O. Avramenko, V. Naradovyi]{Olga Avramenko, Volodymyr Naradovyi}

\address{Vytautas Magnus University, K. Donelaičio g. 58, Kaunas, 44248, Lithuania}
\email{olga.avramenko@vdu.lt}

\address{Volodymyr Vynnychenko Central Ukraine State University, Shevchenko St. 1, Kropyvnytskyi, 25006, Ukraine}
\email{v.v.naradovyi@cuspu.edu.ua}

\subjclass[2020]{76B15, 76B55, 76E99}
\keywords{modulational instability, internal waves, surface tension}

\begin{document}

\begin{abstract}
The propagation of internal waves in a hydrodynamic system comprising a solid bottom and an upper half-space is investigated. The study is conducted within the framework of a nonlinear low-dimensional model incorporating surface tension on an interface using the method of multi-scale expansions. The evolution equation of the envelope of the wave packet takes the form of the Schr\"{o}dinger equation. Conditions for the Benjamin--Feir stability of the solution of the evolution equation are identified for various physical and geometrical characteristics of the system.
\end{abstract}

\maketitle

\section{Introduction}\label{s:1}

The study focuses on examining the modulational instability or Benjamin--Feir instability of wave packets along the interface of liquid layer and half-space above it, considering surface tension. For the first time, modulational instability was presented by Benjamin and Feir  \cite{Benjamin1967}. Zakharov continued to investigate this phenomenon and derived the evolution equation for the envelope of the wave packet in the form of a nonlinear Schr\"{o}dinger equation (NLS) \cite{Zakharov1968}.
In subsequent research, more intricate models have been developed. For example, a higher-order NLS equation featuring fifth-order nonlinearity for wave envelopes on a finite-depth fluid surface was derived in \cite{Sedletsky21}. Additionally, the six-wave interaction and classical three-wave equations were formulated from the free surface gravity wave equation incorporating surface tension in \cite{Ablowitz23}.
Furthermore, noteworthy contributions include the studies \cite{Ionescu2018} and \cite{Düll2021}, which provide analytical insights into the role of surface tension in wave propagation on free surfaces and at fluid interfaces.

Here, we review several studies on the propagation of nonlinear waves in layered fluids examined through the framework of multiple-scale expansions. Hasimoto and Ono \cite{Hasimoto1972} employed this method to describe a weakly nonlinear solution as a modulated wave packet propagating along a water layer, with its envelope governed by the NLS equation.
In \cite{Nayfeh1976}, a detailed analysis of wave packet propagation along the interface of two semi-infinite fluids with surface tension is presented, where the evolution of the envelope is derived as an NLS equation. Grimshaw and Pullin \cite{Grimshaw85} further discussed the stability of finite-amplitude interfacial progressive waves in a two-layer fluid against small perturbations, where an NLS equation is obtained to describe slowly modulated waves using the multiscale expansion.
The study by Christodoulides and Dias \cite{Christodoulides1995} examines both transverse and longitudinal modulations in a fluid layer bounded by a rigid bottom and lid, including the limiting cases of the `lower half-space –- upper layer with rigid lid' and the `half-space -– half-space' configurations.
The nonlinear problem of wave-packet propagation along the interface of two semi-infinite fluids is addressed by Selezov et al. \cite{Selezov03} using a fourth-order multiple-scale method. We also highlight several studies where the multiscale expansion method has been applied to explore nonlinear wave phenomena in two-component hydrodynamic systems with flows \cite{Abrashkin18, Li19, Pal24}.
It is worth noting that the multiple-scale expansion method involves complex transformations, posing challenges as model complexity grows; however, modern computer algebra systems have made it feasible to handle these complexities, which is a central reason for applying this method in the present study.

The following studies investigate the role of the Benjamin--Feir instability within hydrodynamic settings, with a particular focus on its stabilization through dissipation and its impact on extreme wave formation. In \cite{SEGUR2005}, it was demonstrated that for waves with a narrow spectral bandwidth and moderate amplitude, specific types of dissipation can stabilize the instability, a finding that was experimentally validated. Direct numerical simulations in \cite{Wu2006} further support these results, emphasizing the importance of incorporating dissipation models.
Onorato et al. \cite{Onorato2006} reveal that the Benjamin--Feir index is closely related to modulational instability and the probability of extreme wave events, while \cite{ZAKHAROV2009540} and \cite{El2016} discuss the emergence of nonlinear processes and dispersive shock waves driven by modulation instability. Specifically, Zakharov and Ostrovsky \cite{ZAKHAROV2009540} examine nonlinear dynamics arising from modulation instability, whereas El and Hoefer \cite{El2016} provide a comprehensive review of dispersive hydrodynamics, with a focus on dispersive shock waves.
In \cite{Armaroli2018}, regimes balancing wind and viscosity effects in wave propagation are identified and validated through experiments. Additionally, spectral methods presented in \cite{Berti2022, Berti2023} offer new perspectives on the eigenvalues associated with the Benjamin--Feir instability, detailing stability characteristics in the vicinity of the Stokes wave.

As noted above, although there has been substantial interest in the phenomenon of modulational stability, there remains a lack of comprehensive research on its application to internal wave packets at an interface with surface tension effects.

This article addresses the modulational stability of a two-layer hydrodynamic system, specifically the `layer with a solid bottom -- half-space' (La-HS) configuration, incorporating surface tension through the method of multiscale expansions, carried out using symbolic computation. The analysis is carried out using the method of multiple scales within the framework of a weakly nonlinear model.
Building on prior analyses of the `half-space -- half-space' (HS-HS) system \cite{Nayfeh1976}, the present study offers a refined perspective and complements the broader framework introduced in \cite{Christodoulides1995}, where the La-HS configuration was not examined in detail.
The accuracy of the results is validated by examining the limiting cases where the layer thickness becomes infinitely large, effectively transitioning to the HS-HS system.

\section{Problem Statement and Research Method}\label{s:2}

\begin{figure}[t]
	\includegraphics[width=4cm]{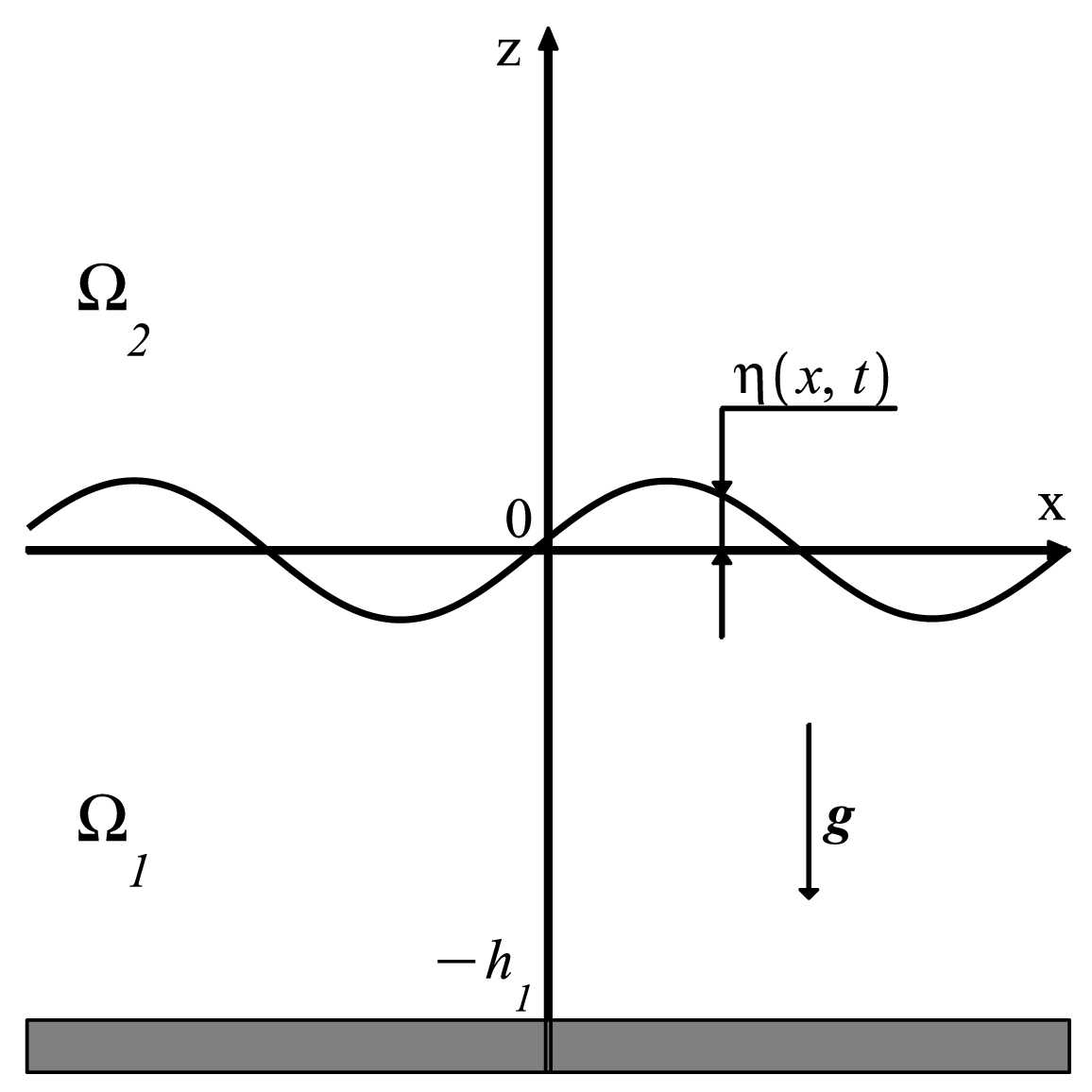}
	\caption{Statement of the problem.}
	\label{fig:Fig1}
\end{figure}

We examine the propagation of wave packets along the interface \( z = \eta(x, t) \) between two incompressible fluid media, \( \Omega_{1} \) and \( \Omega_{2} \), with densities \( \rho_{1} \) and \( \rho_{2} \), respectively, taking into account the surface tension \( T \) acting on the interface \( \eta(x, t) \) (see figure~\ref{fig:Fig1}).
The regions in an undisturbed state have the following form:
$\Omega_{1} = \{(x,z): |x| < +\infty, - h_{1} < z < 0\}$ and $\Omega_{2} = \{(x,z): |x| < +\infty, 0 < z < +\infty\}$ where the thickness of the layer is $h_{1}$.

The mathematical formulation is provided in a dimensionless form, with scaling based on gravitational acceleration \( g \), the density of the lower fluid \( \rho_{1} \), and a characteristic surface tension \( T_{0} \). From these parameters, the characteristic length is defined as \( L = \left( T_{0} \rho_{1}^{-1} g^{-1} \right)^{1/2} \), the characteristic time as \( t_{0} = \left( L g^{-1} \right)^{1/2} \), and the characteristic mass as \( m_{0} = \rho_{1} L^{3} \).
Dimensionless quantities, indicated by an asterisk, are defined as follows
\begin{eqnarray}
	& L(x^{*},z^{*}, h_1^*)=(x,z,h_1), \quad \rho_{1} (\rho_{1}^*, \rho_{2}^*) = (\rho_{1}, \rho_{2}), \quad t_{0} t^{*} = t, \nonumber \\
	& \quad T_{0} T^{*} = T, \quad  \alpha L \eta^*= \eta, \quad \alpha L^{2} t^{-1}_{0} (\phi^{*}_{1}, \phi^{*}_{2})=(\phi_{1}, \phi_{2}). \label{dimensionless}
\end{eqnarray}
where $\alpha = a / l$ is a small dimensionless parameter representing the wave steepness, with $a$ being the maximum displacement of the interface $\eta(x,t)$ and $l$ the wavelength.
It is worth noting that this nondimensionalization approach allows for the exploration of surface tension effects, denoted by $T$, by setting a fixed characteristic value $T_0$.

The propagation velocities of the wave packets in the regions $\Omega_{j}$ are defined in terms of the gradients of the potentials $\phi_{j}(x, z, t)$ for $j=1,2$. Consequently, the mathematical formulation of the wave packet propagation problem within this model takes the form
\begin{eqnarray}
	& \Delta \phi_{j} = 0 \quad \text{in} \quad \Omega_{j} ,  \nonumber\\
	& \eta_{,t} - \phi_{j,z} = - \alpha \eta_{,x} \phi_{j,x} \quad \text{at} \quad z=\alpha \eta(x,t), \label{statmentOfPr}\\
	&  \phi_{1,t} - \rho \phi_{2,t} + (1 - \rho) \eta + 0.5 \alpha \left( \nabla \phi_{1} \right)^{2} - 0.5 \alpha \rho \left( \nabla \phi_{2} \right)^{2} \nonumber\\
	& - T \left( 1 + \left( \alpha \eta_{,x} \right)^{2} \right)^{-1.5}
 \eta_{,xx} = 0 \quad \text{at} \quad z= \alpha \eta(x,t), \nonumber \\
	& \phi_{1,z} = 0 \quad \text{at} \quad z = -h_{1},  \quad
	\left| \nabla \phi_{2} \right| \rightarrow 0 \quad \text{at} \quad z \rightarrow + \infty \nonumber
\label{boundCondLaHS},
\end{eqnarray}
where $\rho= \rho_2/\rho_1$ is the density ratio.

The surface elevation and velocity potentials in the domains $\Omega_{j} (j=1,2)$ are represented according to the method of multiple scales
\begin{eqnarray}
	\eta(x,t) & = & \sum\nolimits_{n=1}^{3} \alpha^{n-1} \eta_{n}(x_{0},x_{1},x_{2},t_{0},t_{1},t_{2}) + \mathrm{O}(\alpha^{3}), \label{multiScaleApr} \\
	\phi_{j}(x,z,t) & = & \sum\nolimits_{n=1}^{3} \alpha^{n-1} \phi_{jn}(x_{0},x_{1},x_{2},z,t_{0},t_{1},t_{2}) + \mathrm{O}(\alpha^{3}), \nonumber
\end{eqnarray}
where $x_{n} = \alpha^{n} x$, $t_{n} = \alpha^{n} t$ are the spatial and temporal scaling variables.

The substitution (\ref{multiScaleApr}) into the problem (\ref{statmentOfPr}) leads to the first three linear approximations with respect to the unknown functions, which are coefficients in the expansion (\ref{multiScaleApr}).

Here are the solutions for the first approximation
\begin{eqnarray}
	&\eta_{1}  =  A \exp(\mathrm{i} \theta) + \overline{A} \exp(-\mathrm{i} \theta),\label{eta1LaHS}\\
	&\phi_{11} = -\frac{\mathrm{i} \omega}{k \sinh(k h_{1})} \left( A \exp(\mathrm{i} \theta ) - \overline{A} \exp(-\mathrm{i} \theta )\right) \cosh(k(h_{1}+z)) , \label{phi11LaHS} \\
	&\phi_{21} = \frac{\mathrm{i} \omega}{k} \left( A \exp(\mathrm{i} \theta - k z) - \overline{A} \exp(-\mathrm{i} \theta - k z )\right), \label{phi21LaHS}
\end{eqnarray}
where $A=A(x_{1},x_{2},t_{1},t_{2})$ is the envelope of the wave packet, $\overline{A}$ is the complex conjugate of $A$, $k$ is the wave number, $\omega$ is the frequency of the wave packet center, $\theta = k x_{0} - \omega t_{0}$.
The dispersion relation is given by
\begin{equation}
	\omega^{2} = \frac{k - \rho k + T k^{3}}{\coth(k h_{1}) + \rho }. \label{dispLaHS}
\end{equation}

Based on these solutions
of the first approximation
(\ref{eta1LaHS})-(\ref{phi21LaHS}) and the dispersion relation (\ref{dispLaHS}), conditions of solvability and solutions of the second approximation are obtained. Below is the analytical expression for $\eta_{2}(x,t)$
\begin{equation}
	\eta_{2} = b_{0} A \overline{A} + \Lambda A^{2} \exp(2 \mathrm{i} \theta) + \mathrm{c.c.} \label{eta2}
\end{equation}
where the coefficients $b_{0}$ and $\Lambda$ have the form
\begin{eqnarray}
	& b_{0} = - \frac{ \omega^{2} }{2 (1 - \rho) \sinh^{2} (k h_{1})}, \nonumber\\
	& \Lambda =  \frac{k \omega^{2} \coth(k h_{1}) \left( 2 \sinh^{2}(k h_{1})(1 - \rho) + 3 \right)  }{2 \cosh(2 k h_{1}) \omega^{2} - \left( 4 T k^{3} + k - \rho - 2 \omega^{2} \rho \right) \sinh(2 k h_{1})}. \nonumber
\end{eqnarray}
The solvability condition for the second approximation is in the form
\begin{equation}
	W_{11} A_{,t_{1}} + W_{12} A_{,x_{1}} = 0, \label{condSolSecond}
\end{equation}
where the coefficients $W_{11}$ and $W_{12}$ depend only on the geometric and physical parameters of the system $T,\rho,k$, and $h_{1}$. After transformations, the condition (\ref{condSolSecond}) can be rewritten as\begin{equation}
	A_{,t_{1}} + \omega^{\prime} A_{,x_{1}} = 0, \label{CSL2}
\end{equation}
where $\omega^{\prime} = \partial \omega / \partial k$ is the group velocity.

For the problem of the third approximation, the solvability condition was found to be
\begin{equation}
	W_{21} A_{,t_{2}} + W_{22} A_{,x_{2}} + W_{23} A_{,x_{1}x_{1}} + W_{24} A^{2} \overline{A} = 0, \label{condSolThird}
\end{equation}
where the coefficients $W_{21}$, $W_{22}$, $W_{23}$, and $W_{24}$ depend only on $T,\rho,k$, and $h_{1}$. After analytical transformations, considering the dispersion relation (\ref{dispLaHS}), equations (\ref{CSL2}) and (\ref{condSolThird}), and transitioning from the scaling variables to variables $x$ and $t$, the evolution equation for the envelope of wave packets on the contact surface can be written as
\begin{equation}
	\mathrm{i} A_{,t} + \mathrm{i} \omega^{\prime} A_{,x} + 0.5 \omega^{\prime \prime} A_{,xx} = - \alpha^{2} \omega^{-1} J A^{2} \overline{A}, \label{nonEqShr1}
\end{equation}
here $\omega^{\prime \prime} = \partial^{2} \omega / \partial k^{2}$ and
\begin{align*} J=\frac{W_1\omega^6+W_2\omega^4+W_3\omega^2+W_4}{W_5\omega^2+W_6} k,
\end{align*}
where
$\qquad	W_1 = \cosh ( 2 k h_{1} ) \mathrm{csch}^{3} (k h_{1} )+ 2 \rho \mathrm{csch} (k h_{1} ) \coth (k h_{1} ) ,
$
\begin{align*}
	& W_2 = 6 k \rho^{3} \sinh^{2} (k h_{1} ) \cosh (k h_{1} )- 2 k \rho^{2} \big[  \cosh ( 2 k h_{1} ) \mathrm{csch} (k h_{1} ) \\
	& + 2 \cosh ( 2 k h_{1} ) \cosh (k h_{1} ) - 6 \sinh (k h_{1} ) \cosh^{2} (k h_{1} ) + \cosh^{3} (k h_{1} )\\
	& + \cosh ( 2 k h_{1} ) \cosh (k h_{1} ) \coth (k h_{1} ) \big ] + k \rho \big[ 2 \cosh ( 2 k h_{1} ) \mathrm{csch} (k h_{1} ) \\
	& +  6 \cosh ( 2 k h_{1} ) \cosh (k h_{1} ) +  1.5 \coth (k h_{1} ) \mathrm{csch} (k h_{1} ) \\
	& - 2  \sinh^{2} (k h_{1} ) \cosh (k h_{1} ) + 2 \cosh ( 2 k h_{1} ) \coth (k h_{1} ) \mathrm{csch} (k h_{1} )\\
	& - 12 \sinh (k h_{1} ) \cosh^{2} (k h_{1} ) + 2 \cosh ( 2 k h_{1} ) \cosh (k h_{1} ) \coth (k h_{1} ) \big ] \\
	& - k \big [ 5.5 \coth (k h_{1} ) \mathrm{csch} (k h_{1} ) + 2 \cosh ( 2 k h_{1} ) \cosh (k h_{1} ) \coth^{2} (k h_{1} ) \\
	& - 6  \cosh^{2} (k h_{1} ) \coth (k h_{1} ) \mathrm{csch} (k h_{1} ) + 2 \coth (k h_{1} ) \mathrm{csch} (k h_{1} ) \cosh^{4} (k h_{1} ) \big ]\\
	& - 4 T k^{3} \mathrm{csch} (k h_{1} ) \coth (k h_{1} ),
\end{align*}
\begin{align*}
	& W_3 = k^2 \big ( 1 - \rho \big ) \bigg[ \big ( 1 - \rho \big ) \{ 2 \rho \cosh (k h_{1} ) \sinh^{2} (k h_{1} )\\
	&  + 6 \sinh (k h_{1} ) \cosh^{2} (k h_{1} )
	- 2 \cosh (k h_{1} ) \cosh ( 2 k h_{1} ) \coth (k h_{1} ) \} \\
	& - 1.5 T k^{2} \mathrm{csch} (k h_{1} ) \cosh ( 2 k h_{1} )
	+ 11 T k^{2} \rho  \cosh (k h_{1} ) \sinh^{2} (k h_{1} )  \\
	& + 24 T k^{2} \sinh (k h_{1} ) \cosh^{2} (k h_{1} )
	- 6.5 T k^{2} \mathrm{csch} (k h_{1} ) \cosh ( 2 k h_{1} ) \cosh^{2} (k h_{1} )\bigg],
\end{align*}
\begin{align*}
	& W_4 = T k^{5} \big ( 1 - \rho \big )  \big[  1.5 \rho -1.5 -6 T k^{2} \big] \cosh (k h_{1} ) \sinh^{2} (k h_{1} ),
	\\
	& W_5 =2 \big ( 1 - \rho \big )\big ( \rho \sinh (k h_{1} ) + \cosh (k h_{1} ) \big ) \big ( \rho \sinh ( 2 k h_{1} ) + \cosh ( 2 k h_{1} ) \big ),
	\\
	& W_6 =  k \big ( 1 - \rho \big ) \big ( \rho - 1 - 4 T k^{2} \big ) \big (  \cosh (k h_{1} ) + \rho \sinh (k h_{1} ) \big ) \sinh ( 2 k h_{1} ).
\end{align*}

Transitioning to a moving frame with the group velocity by substituting $\xi = x - \omega^{\prime} t$ and $\zeta = t$, let's rewrite the envelope equation (\ref{nonEqShr1}) in the form of a NLS
\begin{equation}
	\mathrm{i} A_{,\zeta} +0.5 \omega^{\prime \prime} A_{,\xi \xi} = - \alpha^{2} \omega^{-1} J A^{2} \overline{A} \label{NEShrod}.
\end{equation}

To derive the modulation stability condition of wave packets, let's consider one of the solutions of the equation (\ref{NEShrod}) which depends only on time
\begin{equation}
	A = a \exp(\mathrm{i} \alpha^{2} a^{2} \omega^{-1} J \zeta), \label{solNES}
\end{equation}
where $a$ is a constant. Substituting (\ref{solNES}) into (\ref{eta1LaHS}) and (\ref{eta2}), taking into account the expansion (\ref{multiScaleApr}), and transitioning from variables $(\xi,\zeta)$ to $(x,t)$, we obtain
\begin{equation}
	\eta(x,t) = 2 a \cos(k x - \skew2 \hat{\omega} t) + 2 \alpha a^{2} \left( b_{0} + \Lambda \cos(2 k x - 2 \skew2 \hat{\omega} t)\right) + \mathrm{O} (\alpha^2), \label{eta}
\end{equation}
where $\skew2 \hat{\omega} = \omega - \alpha^2 a^{2} \omega^{-1} J$.

Following the methodology outlined in \cite{Nayfeh1976}, the nonlinear Schr\"{o}dinger equation is linearized by introducing small perturbations into the solution. Specifically, in solution (\ref{solNES}), we introduce small perturbations \( a_1(\zeta, \xi) \) and \( \beta_1(\zeta, \xi) \) to the amplitude and phase, respectively, and substitute the perturbed solution into the evolution equation (\ref{nonEqShr1}). Retaining only the terms of first-order smallness with respect to the perturbation parameter \( \alpha \), we obtain a linear system of two first-order differential equations in terms of the perturbations \( a_1(\zeta, \xi) \) and \( \beta_1(\zeta, \xi) \). The solvability condition for this system yields a dispersion relation, from which it follows that the perturbations remain bounded for all parameters if the condition
\begin{equation}
	J \omega^{\prime \prime} < 0.\label{CMStLaHS}
\end{equation}
 is satisfied.
The full derivation is omitted here, as it formally coincides with the procedure described in \cite{Nayfeh1976}.

It is important to note that an analytical investigation of the limiting behavior of the parameter pair \( (\rho, k) \), corresponding to the condition \( J = 0 \) as \( k \to \infty \), reveals the existence of two vertical asymptotes. One is located at a small density ratio, \( \rho = \frac{2 - \sqrt{2}}{2 + \sqrt{2}} \simeq 0.1716 \), and the other at a significantly larger value, \( \rho = \frac{2 + \sqrt{2}}{2 - \sqrt{2}} \simeq 5.8275 \) that corresponds to the results obtained earlier in \cite{Christodoulides1995} and \cite{Nayfeh1976}.

It is worth noting that condition (\ref{CMStLaHS}) coincides with the condition obtained for the  HS-HS model of wave propagation at the interface between two semi-infinite regions, as described in \cite{Nayfeh1976}.
To confirm the correctness of the analytical results for the La-HS system, a limiting transition to the HS-HS system was performed as $h_1 \to +\infty$. All equations and expressions for $\Lambda$ and $J$ degenerate into the corresponding equations and expressions previously derived in \cite{Nayfeh1976} for the HS-HS system.
This confirms the correctness of the obtained analytical results.

\section{Modulational stability analysis}

\subsection{Notation overview for stability diagrams}

In this section, a description and analysis of stability diagrams  on the
$(\rho,k)$ plane for La-HS system is presented.
Each stability diagram is divided into regions of linear instability (dark shading) and linear stability. The region of linear stability, in turn, consists of areas of nonlinear stability (unshaded) and nonlinear instability, or in other words - instability of the envelope of the wave packet, or
modulational instability, or Benjamin--Feir instability (light shading).

The condition for linear stability  is determined by the same relationship, namely, when the wave
numbers are greater than a critical value,
$k > k_{c} = \sqrt{(\rho - 1)/T}$, therefore, in all the stability
diagrams presented below, the $(\rho,\ k)$ plane is divided by the
curve $k = k_{c}$ (black curve) into a region of linear instability
located below this curve, and a region of linear stability located above
and to the left of it.

As previously indicated in (\ref{CMStLaHS}), the sign of the expression $J\omega''$
determines whether the envelope of the wave packet is
stable or not, i.e., whether modulational stability exists or not. Thus,
the region of linear stability is divided into regions of nonlinear
stability and instability by curves along which $J = 0$ (red
curves), $J \rightarrow \infty$ (blue curves), and $\omega^{''} = 0$
(green curves).

Below is a description of the regions of modulational
stability and instability, which will be referred to as `stability' and
`instability' regions for brevity in the following discussion.

\subsection{Influence of geometrical parameters on modulational stability}
\subsubsection{A formal description of the stability diagrams}
\begin{figure}[t]
	\begin{minipage}[t]{0.32\linewidth}
		\center{\includegraphics[width=\linewidth]{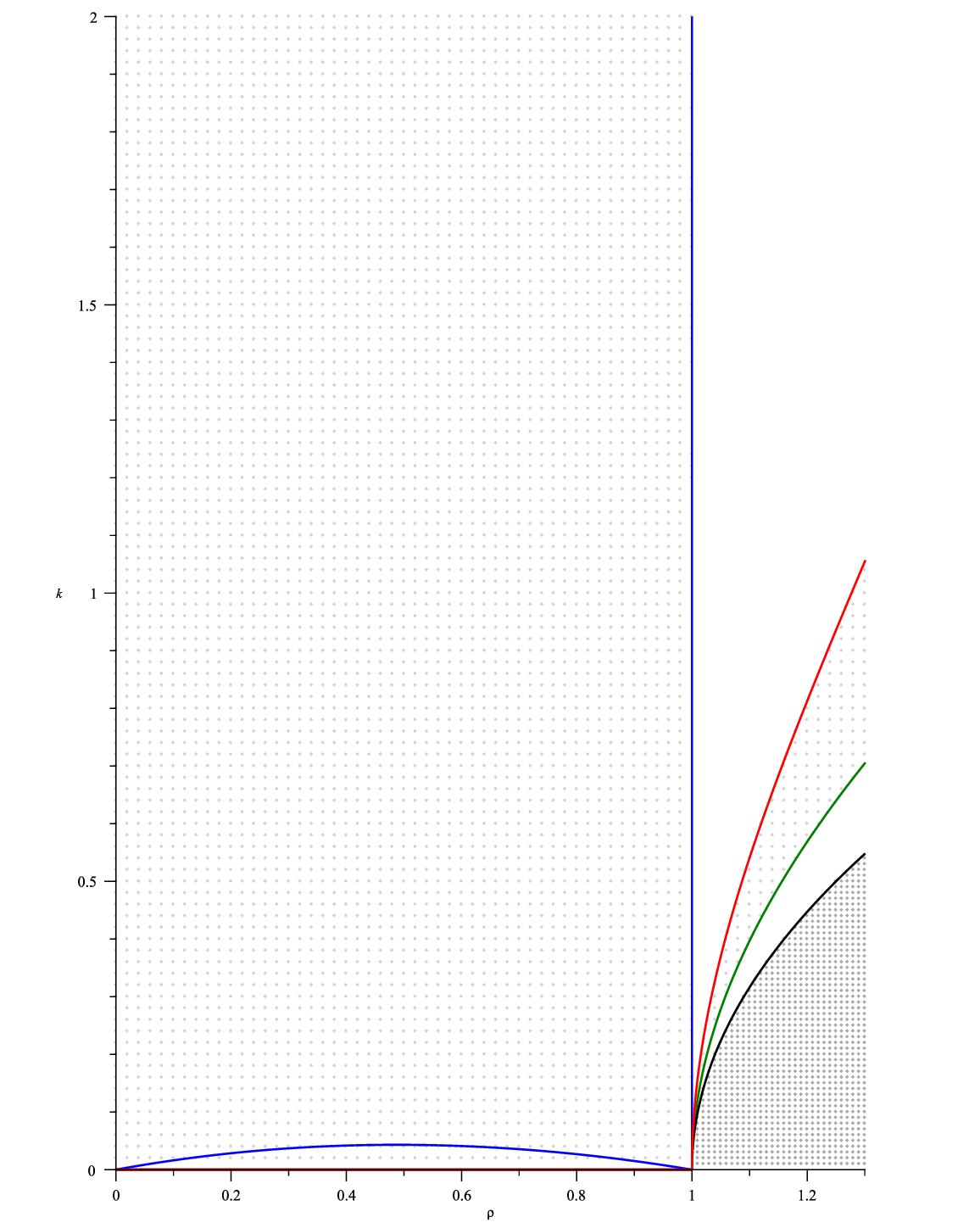}  a) $h_1=0.5$}
		\label{fig:Fig5a}
	\end{minipage}
	\hfill
	\begin{minipage}[t]{0.32\linewidth}
		\center{\includegraphics[width=\linewidth]{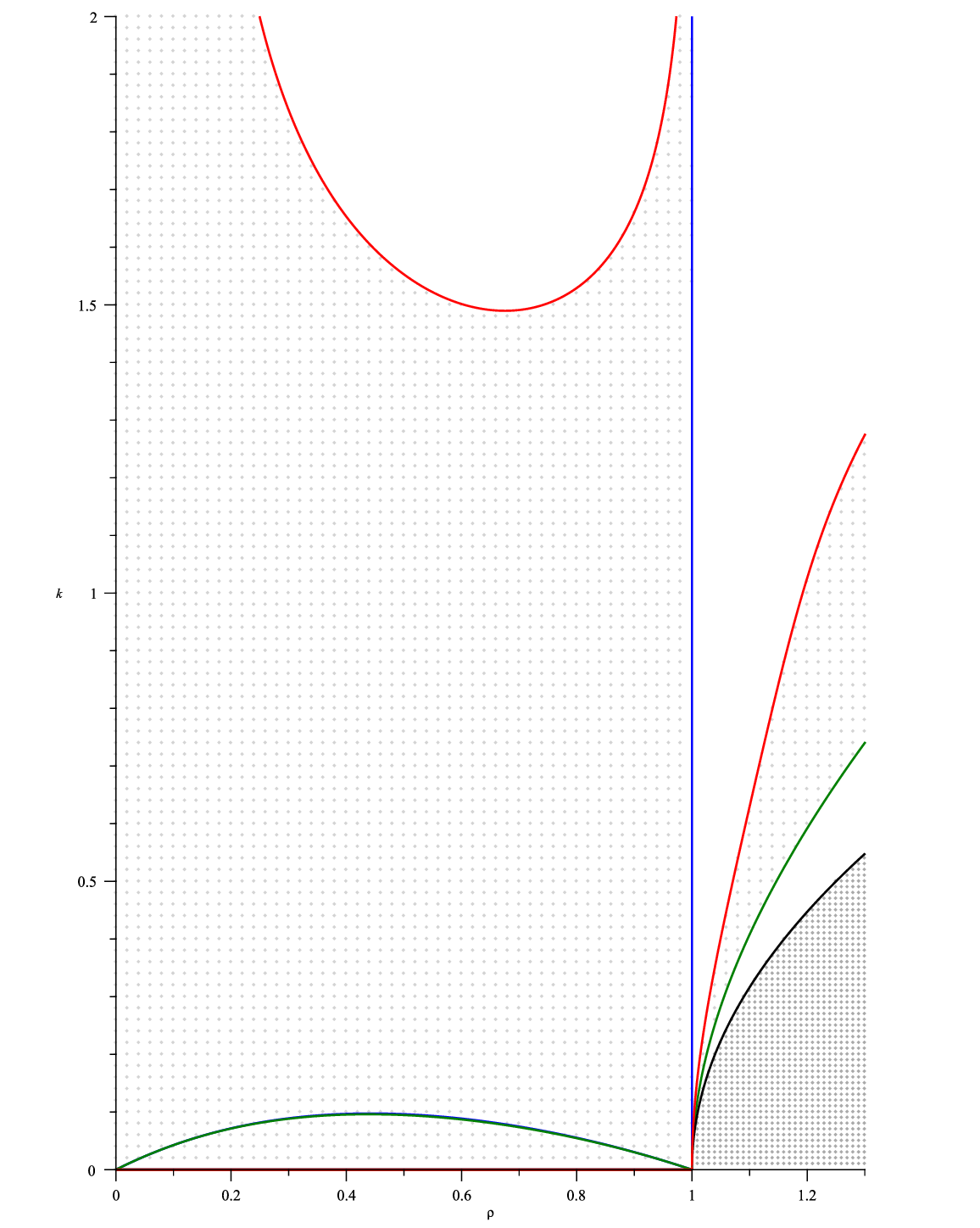}  b) $h_1=1$}
		\label{fig:Fig5b}
	\end{minipage}
	\hfill
	\begin{minipage}[t]{0.32\linewidth}
		\center{\includegraphics[width=\linewidth]{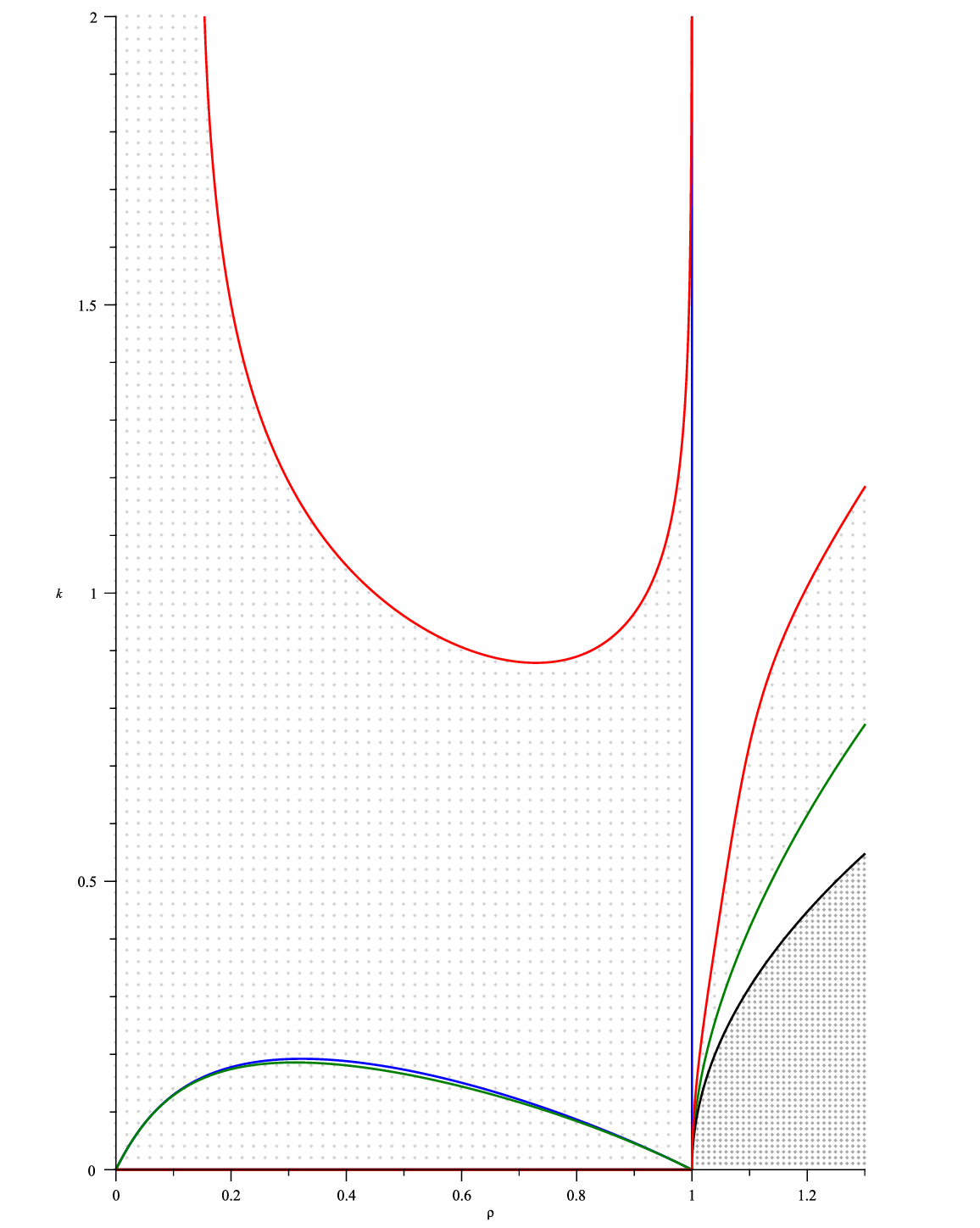} c) $h_1=1.5$}
		\label{fig:Fig5c}
	\end{minipage}
	\vfill
	\begin{minipage}[t]{0.32\linewidth}
		\center{\includegraphics[width=\linewidth]{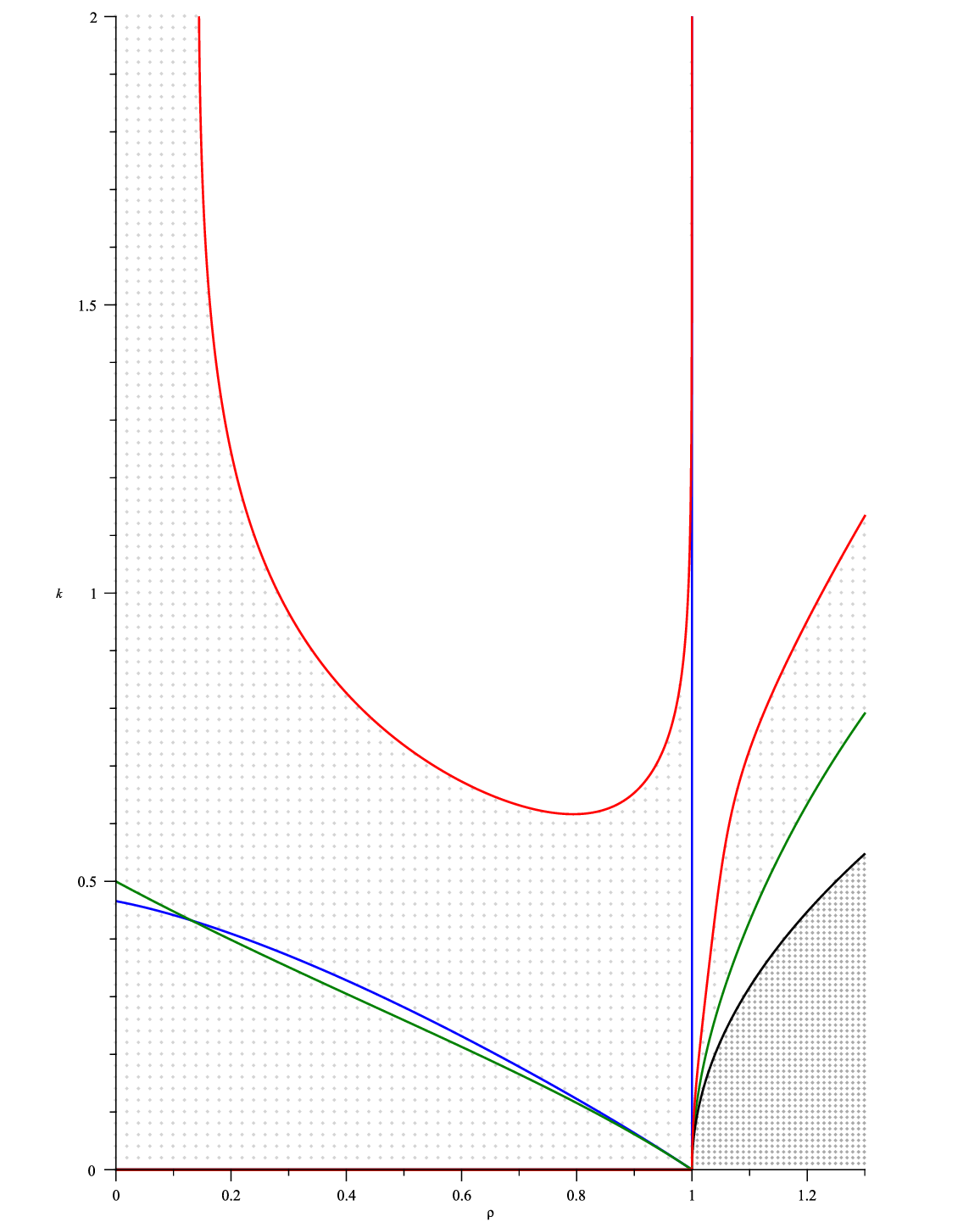}  d) $h_1=2$}
		\label{fig:Fig5d}
	\end{minipage}
	\hfill
	\begin{minipage}[t]{0.32\linewidth}
		\center{\includegraphics[width=\linewidth]{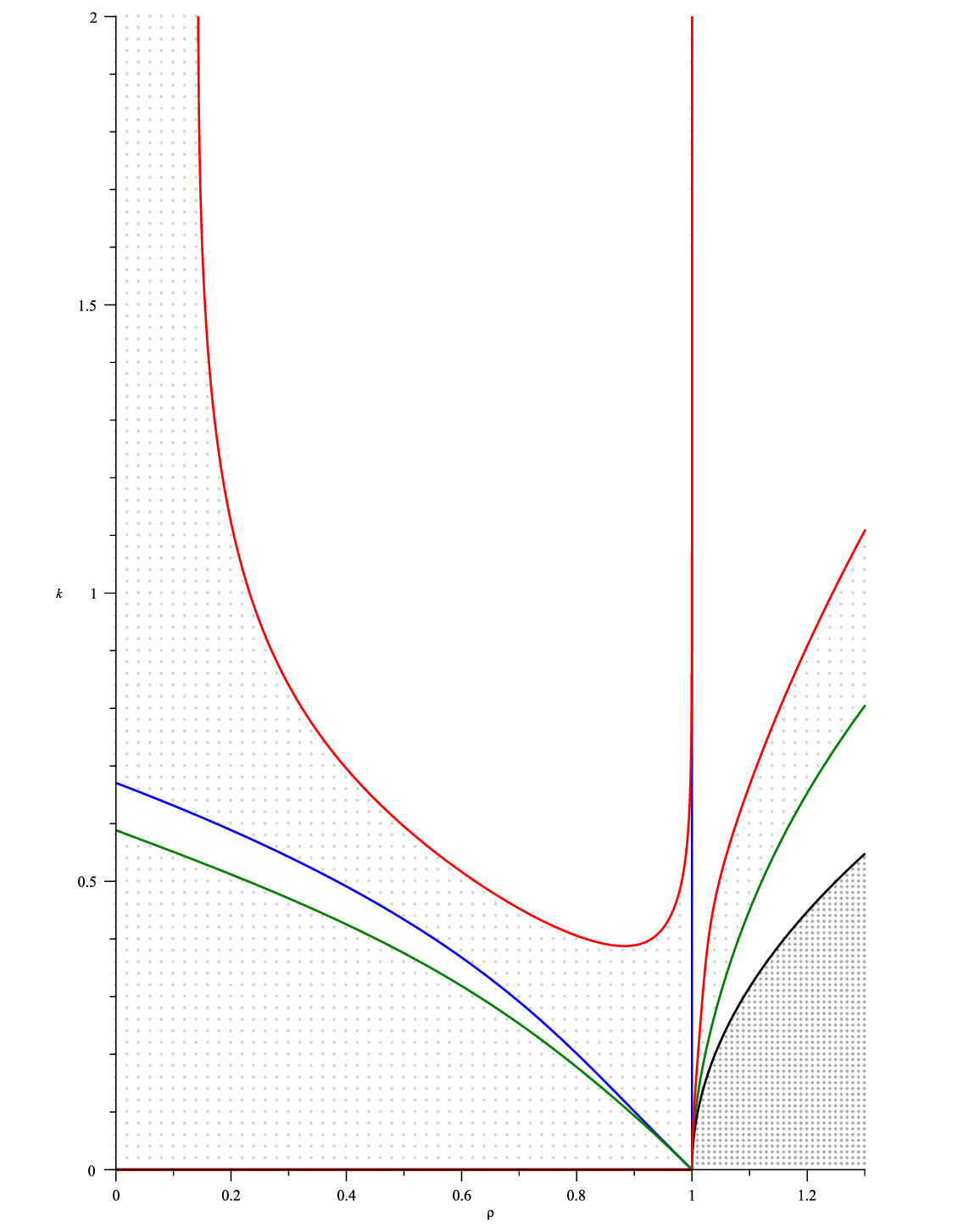}  e) $h_1=3$}
		\label{fig:Fig5e}
	\end{minipage}
	\hfill
	\begin{minipage}[t]{0.32\linewidth}
		\center{\includegraphics[width=\linewidth]{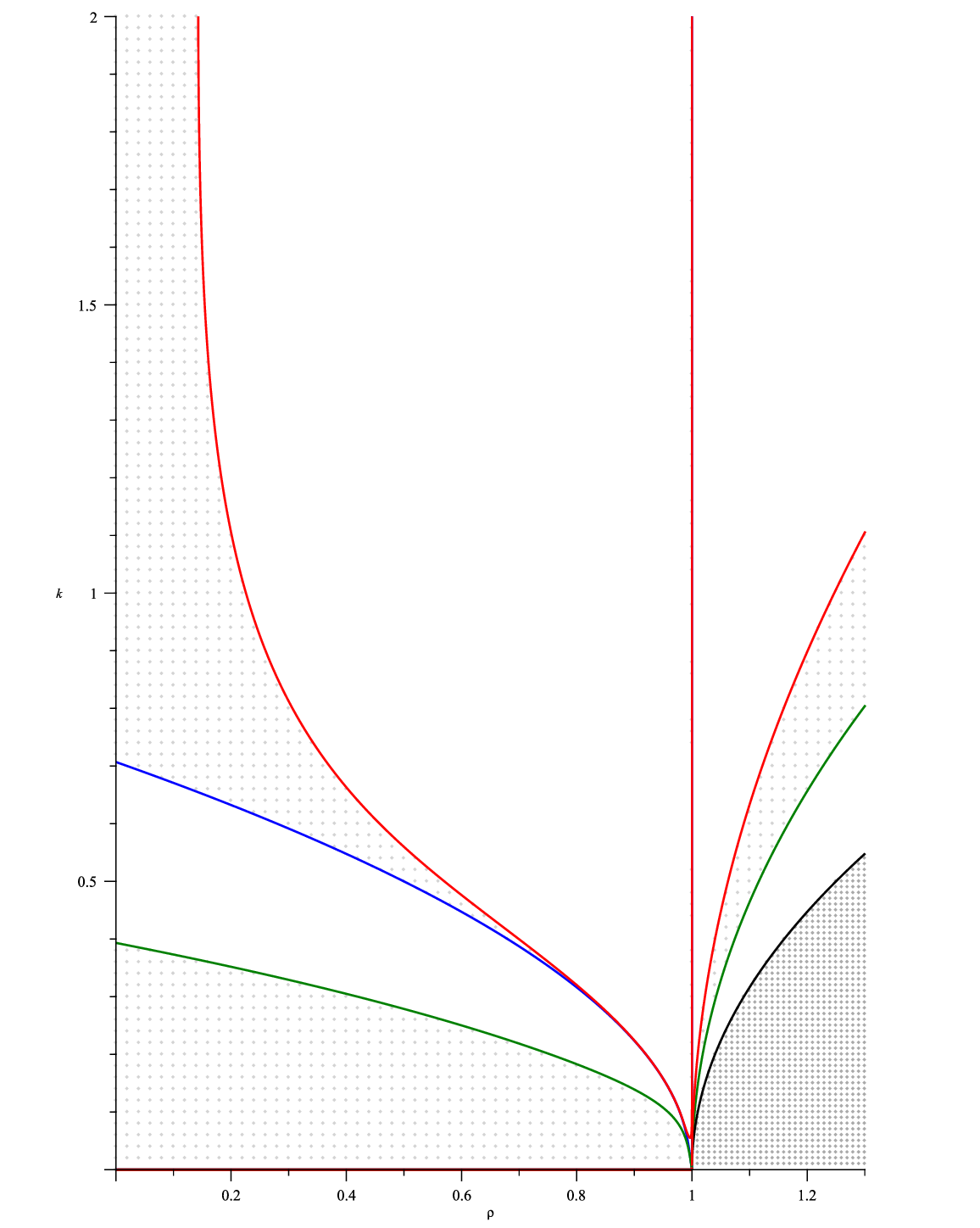} f) $h_1=20$}
		\label{fig:Fig5f}
	\end{minipage}
	\caption{Modulational stability diagram for $T=1$.}
	\label{fig:Fig2}
\end{figure}

The stability diagrams at different layer thicknesses and $T = 1$  are presented in figure~\ref{fig:Fig2}. For all investigated parameter values, there exists a certain similarity in the diagrams  in the right part, where the density ratio $\rho > 1$. In this region, the linear instability
region lies below the curve $k = k_{c}$ (black curve), while above it, three alternating regions of nonlinear stability and instability are situated. The vertical line $\rho = 1$, along which $J \rightarrow \infty$, separates the region of nonlinear stability to the right from the region of nonlinear instability to the left.

We can also mention the presence of a region of nonlinear
stability with asymptotes $\rho = 1$ and $\rho \simeq 0.1716$ for
capillary waves when $\rho < 1$. We will refer to this region as the `upper' region. For a layer thickness of $h_1 = 0.5$, the `upper' region of nonlinear stability begins at $k > 2$, which is why it is not visible in figure~\ref{fig:Fig2}~\textit{a}. As the layer thickness increases to $h_1 = 1, 1.5, 2, 3, 20$ (see figures~\ref{fig:Fig2}~\textit{b, c, d, e, f}), the `upper' region descends, and its lower point tends towards $(1, 0)$ as $h_1 \rightarrow +\infty$.

\begin{figure}[t]
	\includegraphics[width=.3\textwidth]{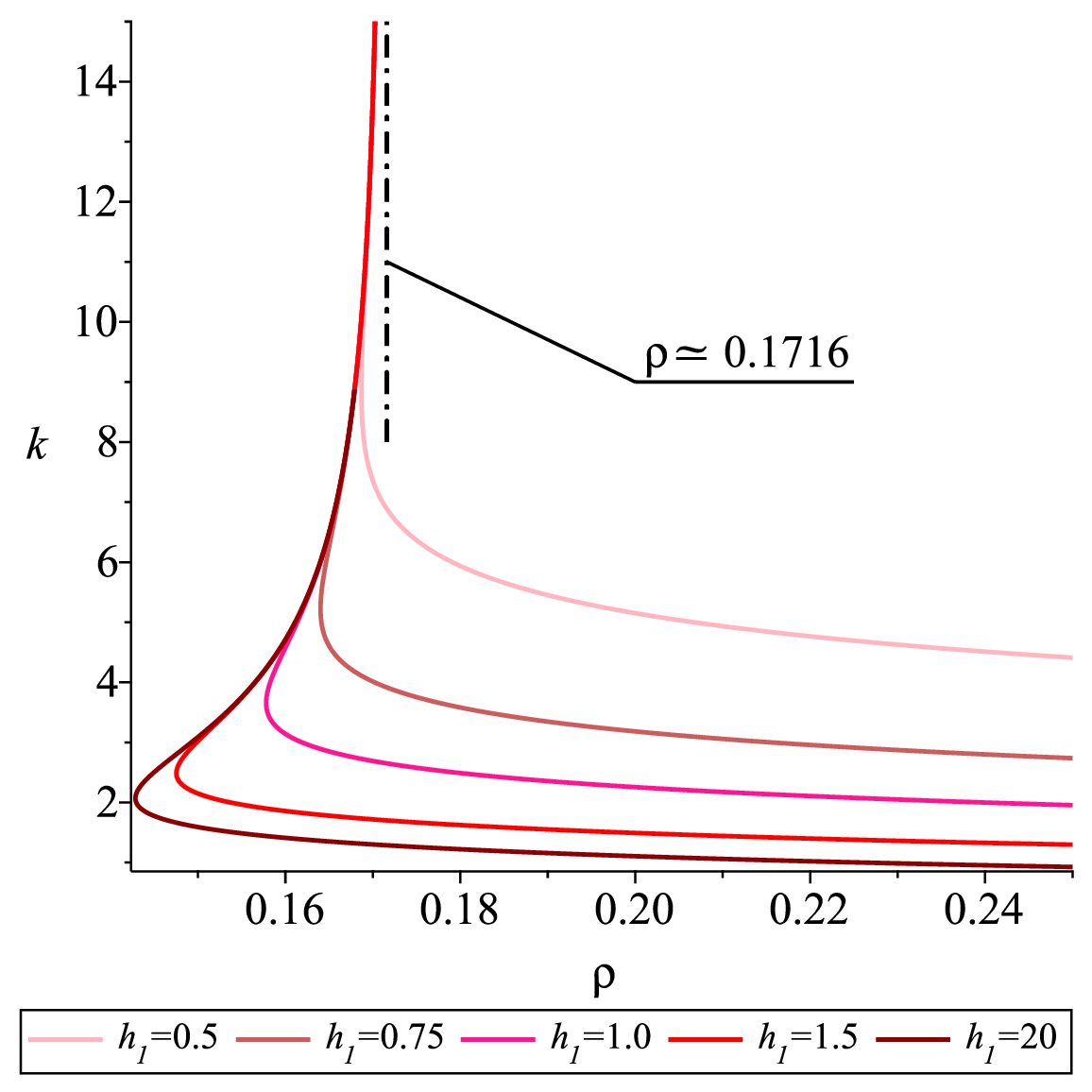}
	\caption{The vicinity of the asymptote for $T=1$.}
	\label{fig:Fig3}
\end{figure}

On figure~\ref{fig:Fig3}, the region around the asymptote $\rho \simeq 0.1716$ is presented for different values of layer thickness
$h_1 = 0.5, 1, 1.5, 2, 3, 20$ at $T = 1$.
It's worth noting that we observe the
following dependence: the thinner the layer, the further to the right
the curve separating the instability region from the `upper' stability
region. The limiting position of these curves as the layer thickness increases
$h_1\rightarrow+\infty$ remains the same, corresponding to the case of
two half-spaces system \cite{Nayfeh1976}, which is evident at $h_1 = 20$.

In the lower part of the diagram, there is an `elongated' region, bounded above by a blue curve $J \rightarrow \infty$  and below by a green curve  $\omega^{''} = 0$.
For small layer thicknesses $h_1$, it essentially degenerates
into a 'cut' on the $(\rho,\ k)$ plane, which is clearly visible for
$h_1 = 0.5, 1, 1.5$ (figures~\ref{fig:Fig2}~\textit{a, b, c}).
 For these values of layer
thickness, the blue curve along which $J \rightarrow \infty$ and the
green curve where $\omega^{''} = 0$ almost coincide creating a `cut' on the plane $(\rho, k)$ which starts from the origin and ends at the point $(1,0)$.

As the layer thickness
increases, both curves rise above the origin, still not diverging far
from each other, having a point of intersection, as seen for $h_1 = 2$
(figure~\ref{fig:Fig2}~\textit{d}). For even larger thicknesses, the `elongated' region of
nonlinear stability expands, as evident for $h_1 = 3, 20$ (figures~\ref{fig:Fig2}~\textit{e, f}). As $h_1 \rightarrow +\infty$, this region completely coincides with
the corresponding region of the two liquid half-spaces system \cite{Nayfeh1976}, and as $h_1 \rightarrow +\infty$ and $\rho = 0$, it corresponds to  \cite{Zakharov1968}.

\subsubsection{Physical interpretation and discussion}
We first note the case of small lower-layer thicknesses (see figures~\ref{fig:Fig2}~\textit{a,~b,~c}). For instance, when the lower layer has a thickness of $h_1=0.5$ (figure~\ref{fig:Fig2}~\textit{a}), the diagram shows that for wave numbers in the range $0<k<2$ and density ratios $\rho<1$, a nearly continuous region of modulational instability emerges. This hydrodynamic configuration leads to a weakening of dispersion, as the wavelength is sufficiently long to be influenced by the finite depth of the lower layer, while the upper half-space does not restrict the propagation of perturbations. Under such conditions, nonlinear effects are enhanced, since weak dispersion cannot effectively suppress them.
Moreover, in this case, there is a deficiency of inertia in the lower layer, while the dominant inertial contribution is concentrated in the upper half-space. This imbalance leads to instability of the interface, as the lower layer is unable to effectively resist deformation.

A special case arises when \( \rho \ll 1 \), indicating that the upper layer is significantly lighter. This amplifies the non-gravitational character of the waves, rendering the interface highly sensitive to small perturbations that may evolve even under weak nonlinearity. In this regime, a substantial restoring force supports oscillations, and modulational instability may occur. This is confirmed by the presence of one of the vertical asymptotes discussed previously, which delineates the instability region for \( \rho < 0.1716 \) observed in all stability diagrams.

Similar phenomena are observed at the bottom of each of the diagrams for \( h_1 = 1, 1.5, 2 \) (figures~\ref{fig:Fig2}~\textit{b,~c,~d}), where the `upper' region of modulational stability emerges for \( k > 1.48 \), \( k > 0.87 \), and \( k > 0.61 \), respectively. As the thickness of the lower layer further increases to \( h_1 = 3, 20 \) (figures~\ref{fig:Fig2}~\textit{e,~f}), in addition to the expansion of the `upper' stability zone, a new `elongated' region of stability appears. This indicates a significant change in the system's behavior due to the increased mass and inertia of the fluid beneath the interface. As a result, dispersion and its stabilizing influence are enhanced. Specifically, with increasing \( h_1 \), the lower layer becomes more actively involved in sustaining wave motion, thereby suppressing modulational instability, as dispersion more effectively competes with nonlinear effects. Thus, increasing \( h_1 \) acts as a stabilizing factor, shifting the boundary between stable and unstable regimes and giving rise to a well-developed region of modulational stability in the \( (\rho, k) \) parameter space.

Let us now consider the case of nearly equal densities of the upper and lower fluids. Evidently, as the density ratio approaches unity (\( \rho \to 1 \)), the difference between the two fluids vanishes, and the interface loses its dynamic significance and wave motion ceases. Physically, this means that for identical densities, no restoring force exists to sustain oscillations of the interface. However, assuming the presence of a thin interfacial layer that generates surface tension \( T \), as in the present study, the diagrams in figure~~\ref{fig:Fig2} demonstrate the existence of a vertical boundary at \( \rho = 1 \) separating regions of instability and stability. For relatively small values of the lower-layer thickness \( h_1 = 0.5, 1.0, 1.5 \) (figures~\ref{fig:Fig2}~\textit{a,~b,~c}), this boundary sharply delineates the domains: instability occurs for \( \rho < 1 \), while stability is observed for \( \rho > 1 \). As the thickness increases, modulational instability is progressively suppressed, which is particularly evident near \( \rho = 1 \) for \( h_1 = 2, 3, 20 \) (figures~\ref{fig:Fig2}~\textit{d,~e,~f}). It is worth noting that in the case of two semi-infinite fluids, as studied by Nayfeh, the left and right regions are not separated by \( \rho = 1 \) but instead merge seamlessly, which corresponds to the limiting case of the present system as \( h_1 \to \infty \). This behavior is well approximated by the near-asymptotic case \( h_1 = 20 \), illustrated in figure~\ref{fig:Fig2}~\textit{f}.

We now turn to a physically and mathematically interesting case in which the upper fluid is denser than the lower one, i.e., the density ratio satisfies \( \rho > 1 \). This configuration represents a linearly unstable system; however, the presence of surface tension at the interface strongly affects the nature of wave dynamics. Specifically, in the case of long-wave perturbations, that is for wave numbers \( k < k_c \), gravitational instability arises due to the heavier fluid overlying the lighter one. This manifests as a region of linear instability, indicated by dark shading in figure~\ref{fig:Fig2}, where small perturbations grow exponentially with time. Such behavior implies that the system is physically unstable in any nonlinear approximation.

As the wave number \( k \) increases, the system transitions out of the regime of linear instability. The diagrams in figure~\ref{fig:Fig2} show results for the range \( 1 < \rho < 1.2 \), where alternating zones of modulational stability and instability are observed. This phenomenon is attributed to the fact that at intermediate and high values of \( k \), surface tension begins to compete with gravitational forces, thereby stabilizing the interface. Nevertheless, nonlinear interactions in certain regimes give rise to modulational instability. The observed alternation of stability and instability zones is a consequence of a delicate balance between dispersion, nonlinearity, and capillary effects.

\begin{figure}[t]
	\includegraphics[width=.3\textwidth]{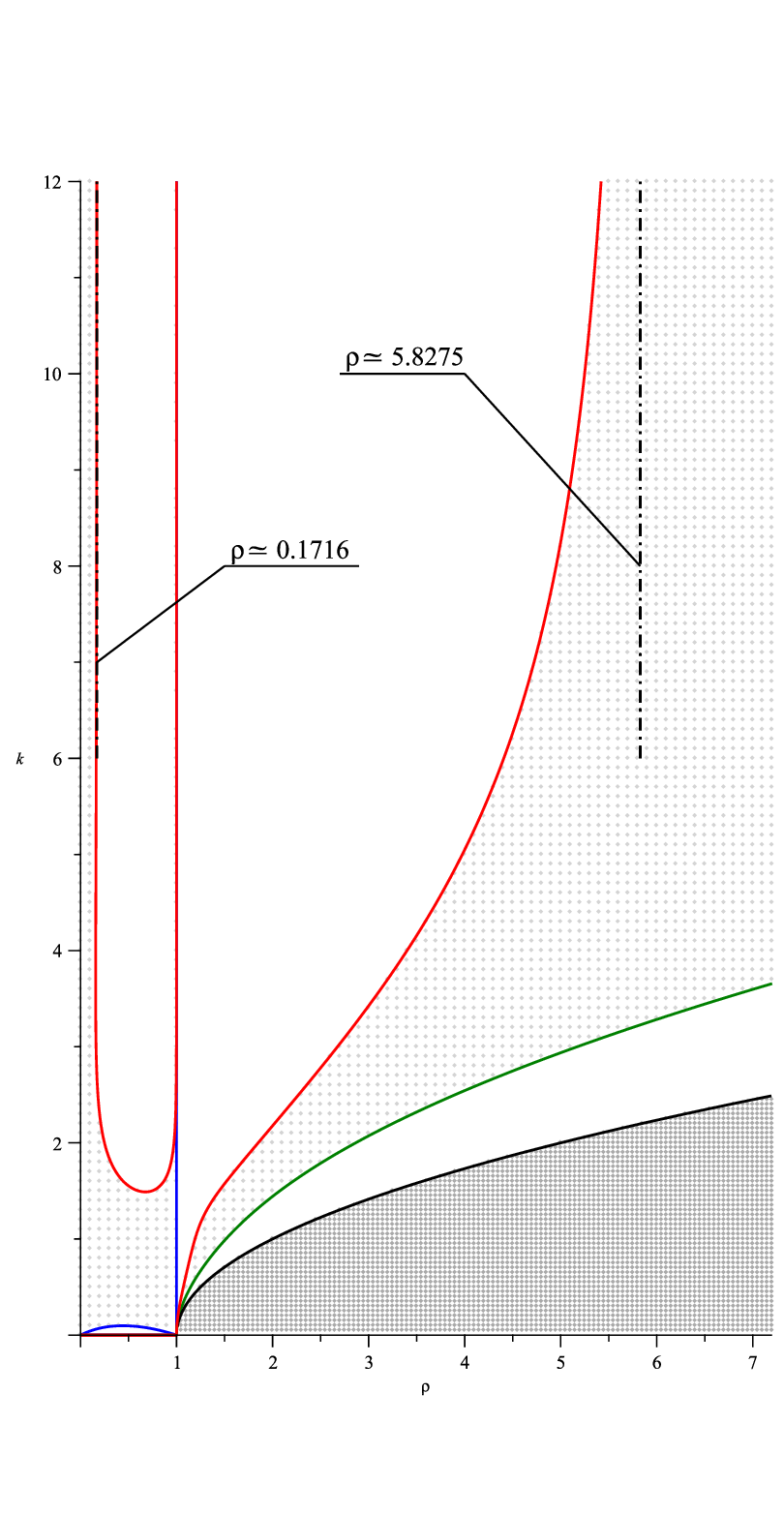}
	\caption{Modulational stability diagram for $h_1=1$ and $T=1$ over the extended range of $\rho$ and $k$.}
	\label{fig:Fig4}
\end{figure}

It should be noted that for density ratio values not represented in the diagrams of figure~\ref{fig:Fig2}, there exists the vertical asymptote mentioned in the previous section, located at \( \rho \simeq 5.8275 \), beyond which lies a region of instability at sufficiently large wave numbers. For clarity, figure~\ref{fig:Fig4} presents the stability diagram for $h_1=1$ and $T=1$ over the extended range \( 0 < \rho < 7.2 \) and \( 0 < k < 12 \), clearly illustrating the presence of both asymptotes.
 Thus, when the critical density \( \rho \simeq 5.8275 \) is exceeded, only a bounded region of modulational stability remains above the zone of linear instability where capillary effects temporarily stabilize wave propagation within a certain range of \( k \) along with an unbounded region of modulational instability, indicating that capillary forces are no longer sufficient to counteract the gravitational dominance of the heavier upper fluid.

\subsection{Influence of surface tension on modulational stability}

\subsubsection{A formal description  of the stability diagrams}

\begin{figure}[t]
	\begin{minipage}[t]{0.29\linewidth}
		\center{\includegraphics[width=\linewidth]{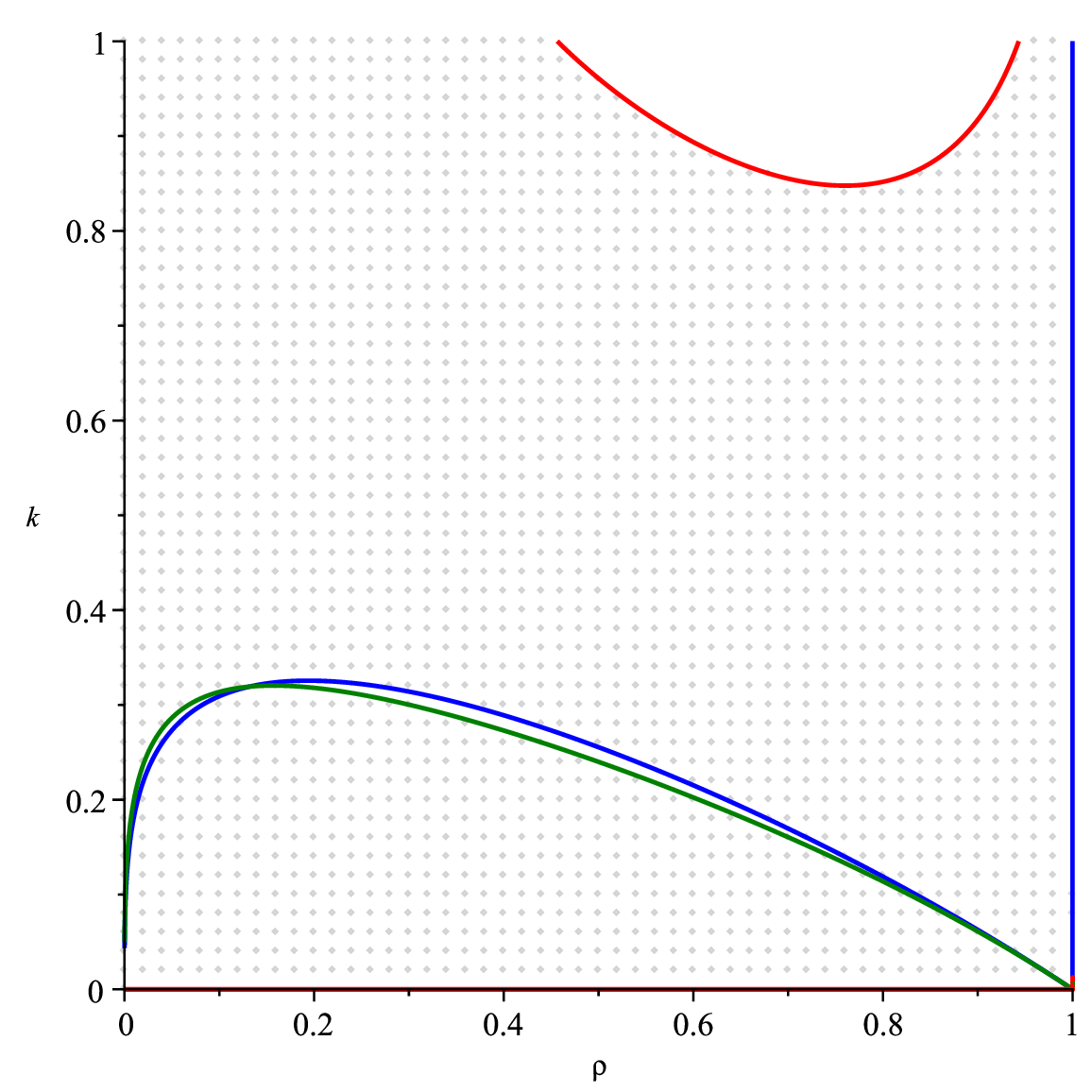}  a) $h_1=1.5 $, $ T=0.75$}
		\label{fig:Fig5a}
	\end{minipage}
	\hfill
	\begin{minipage}[t]{0.29\linewidth}
		\center{\includegraphics[width=\linewidth]{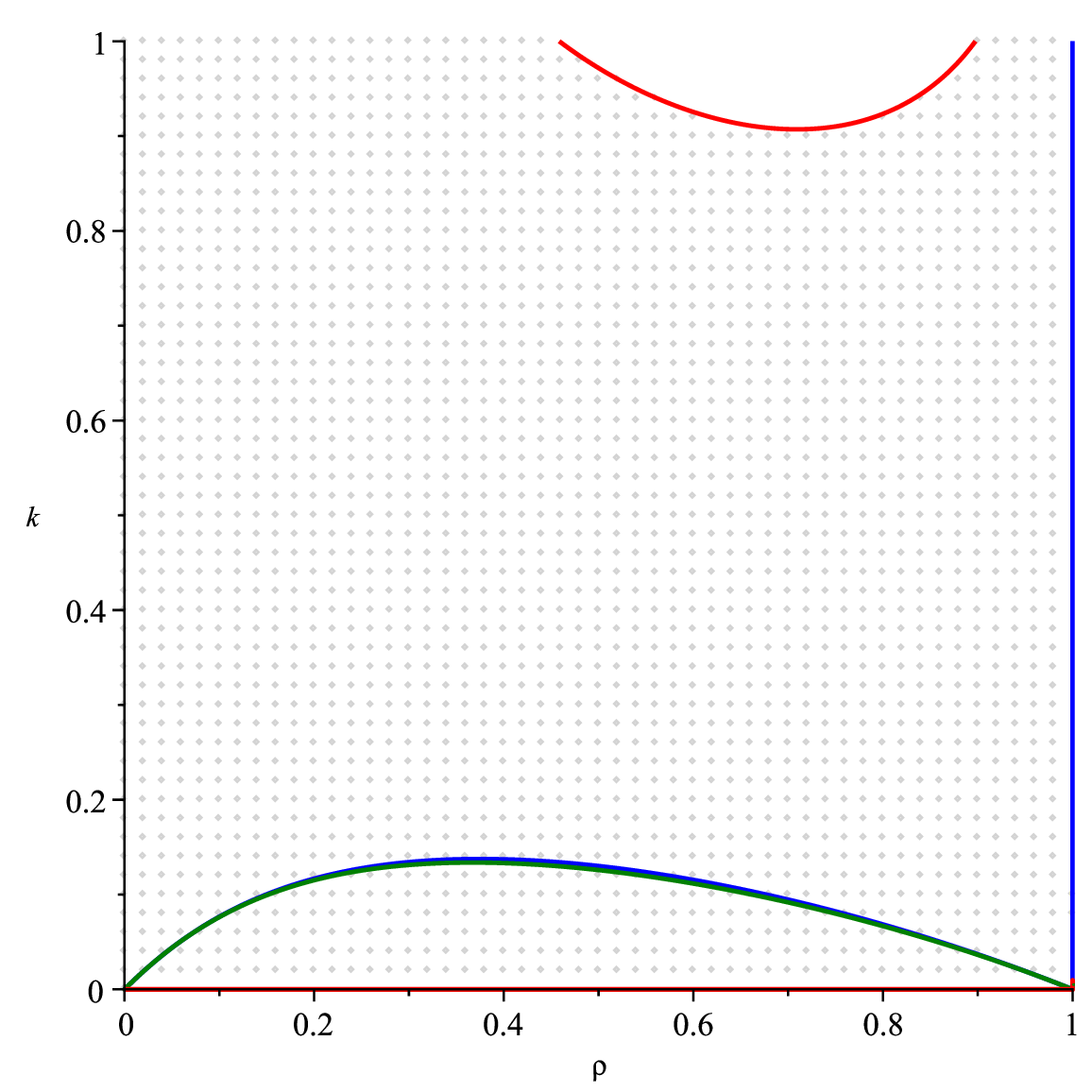}  b)  $h_1=1.5 $, $ T=1.25$}
		\label{fig:Fig5b}
	\end{minipage}
	\hfill
	\begin{minipage}[t]{0.29\linewidth}
		\center{\includegraphics[width=\linewidth]{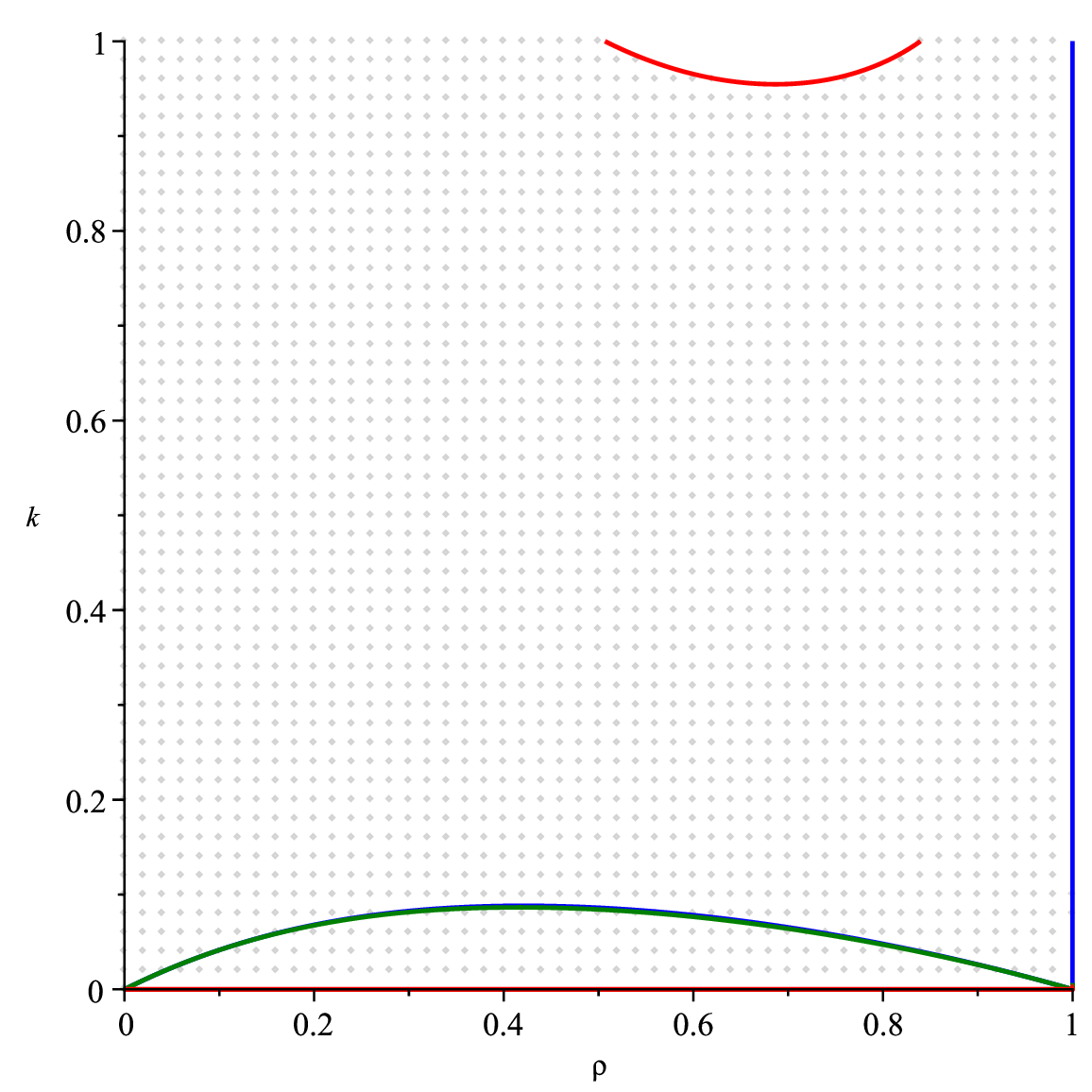} c) $h_1=1.5$, $T=1.75$}
		\label{fig:Fig5c}
	\end{minipage}
	\vfill
	\begin{minipage}[t]{0.29\linewidth}
		\center{\includegraphics[width=\linewidth]{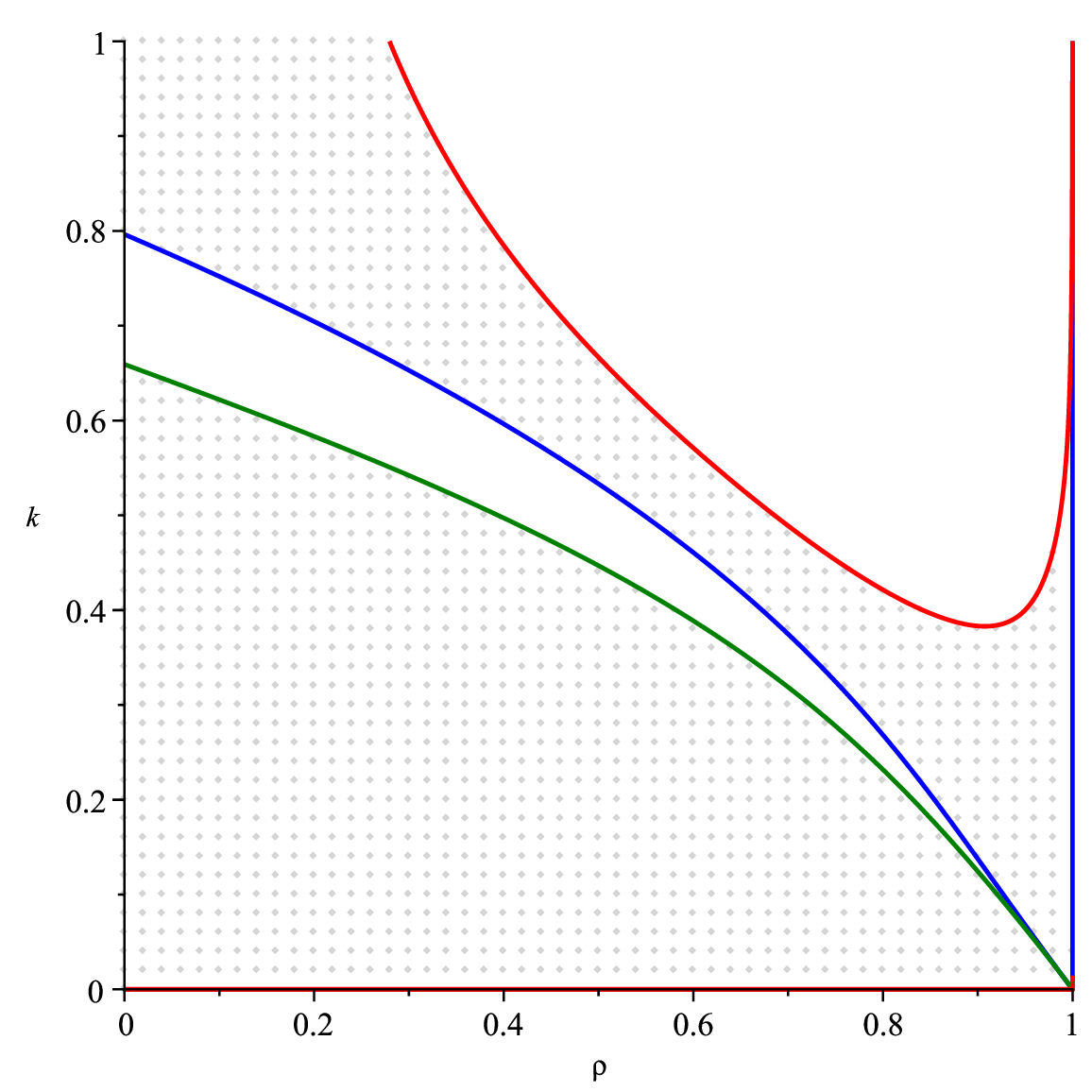}  d) $h_1=3 $, $ T=0.75$}
		\label{fig:Fig5d}
	\end{minipage}
	\hfill
	\begin{minipage}[t]{0.29\linewidth}
		\center{\includegraphics[width=\linewidth]{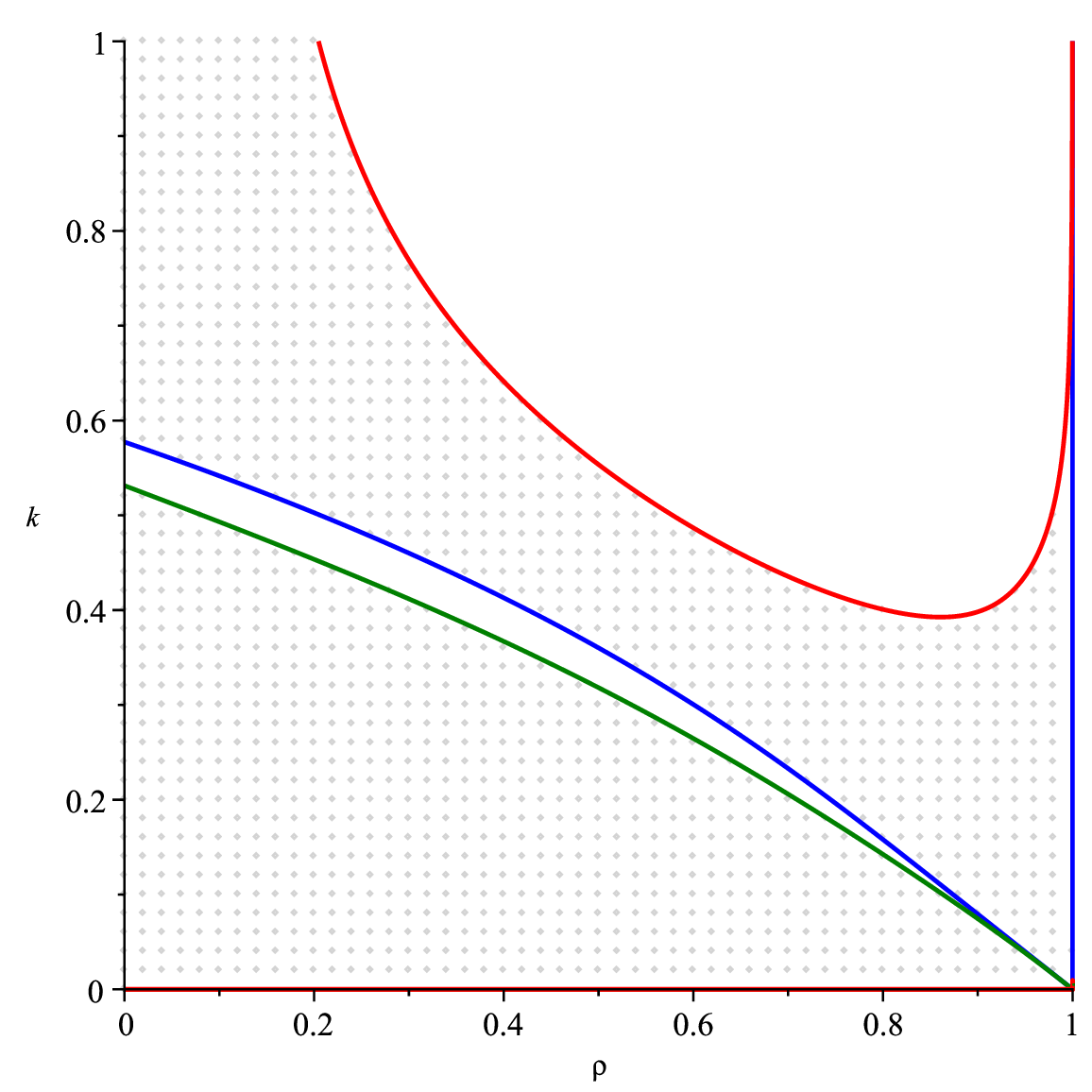}  e) $h_1=3 $, $ T=1.25$}
		\label{fig:Fig5e}
	\end{minipage}
	\hfill
	\begin{minipage}[t]{0.29\linewidth}
		\center{\includegraphics[width=\linewidth]{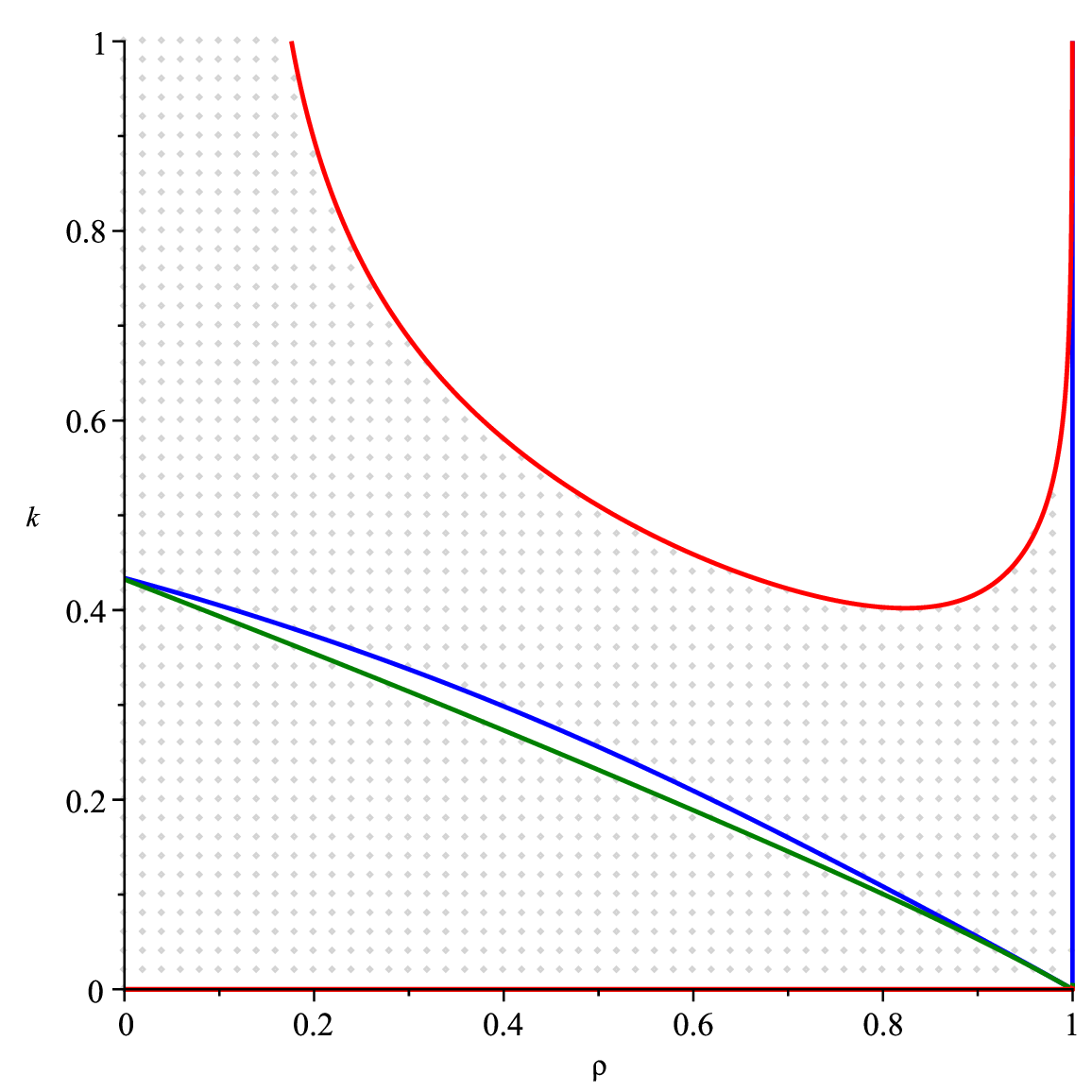} f) $h_1=3$, $ T=1.75$}
		\label{fig:Fig5f}
	\end{minipage}
	\caption{Modulational stability diagrams.}
	\label{fig:Fig5}
\end{figure}

We now present an analysis of the dependence of modulational stability on surface tension. Figures~\ref{fig:Fig5}~\textit{a,~b,~c}
depict stability diagrams for a layer thickness of $h_1 = 1.5$ with
surface tension coefficient values of $T = 0.75, 1.25, 1.75$, while figures~\ref{fig:Fig5}~\textit{d,~e,~f} represent diagrams for $h_1 = 1.5$
with the same values of $T = 0.75, 1.25, 1.75$. It can
be observed that as the surface tension increases, all regions of the
stability diagram undergo significant deformation.
As the surface tension increases $(T = 0.75, 1.25, 1.75)$ with a
layer thickness of $h_1 = 1.5$, the `cut', which is a degeneration of the `elongated' region, in the long-wave region diminishes and approaches the horizontal axis, while the `upper' stability region slightly shifts upward and
widens. Furthermore, when the layer thickness is $h_1 = 3$, the `elongated' stability region significantly narrows as the surface tension increases
$T = 0.75, 1.25$, almost turning into a `cut' in the instability region at $T = 1.75$. Meanwhile, the lower point of the `upper' stability region slightly rises upward, and the region itself becomes substantially wider.

\subsubsection{Physical interpretation and discussion}

The analysis of the diagrams in figure~\ref{fig:Fig5}, corresponding to the case \( \rho < 1 \), demonstrates that surface tension \( T \) significantly affects wave dynamics. In general, increasing \( T \) stabilizes short-wave modes and suppresses the development of modulational instability; however, the effect strongly depends on the geometry of the system. For instance, at a lower layer thickness of \( h_1 = 1.5 \) (figures~\ref{fig:Fig5}~\textit{a,~b,~c}), increasing \( T \) compresses the `cut' region in the lower part of the diagram and slightly elevates the upper stability region, thereby expanding the overall instability zone. For a greater thickness \( h_1 = 3 \) (figures~\ref{fig:Fig5}~\textit{d,~e,~f}), a different behavior is observed: the elongated region of stability at small wave numbers narrows into a thin `cut' as \( T \) increases, while the upper stability region expands. This results in a net reduction of the instability zone. Hence, surface tension plays a complex role in shaping the modulational stability diagram, substantially affecting the distribution of stable and unstable regimes. Depending on the thickness of the lower layer, it may either enhance or suppress modulational instability. These observations underscore the importance of detailed parametric analysis and the necessity of numerical modeling for the quantitative characterization of stability boundaries in various geometric and physical configurations.

\section{Conclusions}

We summarize the key features of modulational stability regions in hydrodynamic system a hydrodynamic system consisting of a solid bottom and an upper half-space on the $ (\rho, k) $ plane, examining various layer thicknesses and surface tension values.

For cases where $ \rho < 1 $, two main stability regions are identified: the `upper' region and the `elongated' region. Specifically, (i) the `upper' region occurs at high wavenumbers, spanning a wide range of density ratios, with boundaries at approximately $ \rho \simeq 0.1716 $ and $ \rho = 1 $; (ii) the `elongated' region, under certain parameter conditions, reduces to a narrow `cut' in the lower part of the stability diagram, nearly vanishing, where wave packet envelopes are unstable across most parameter values for long wavelengths. In this case, for small lower-layer thicknesses, a nearly continuous region of modulational instability emerges, due to weak dispersion and inertia. Increasing the lower-layer depth enhances dispersion and inertial effects, leading to the formation and expansion of stable regions. For near-equal densities (\( \rho \to 1 \)), surface tension becomes the key stabilizing factor, especially for intermediate or large lower-layer depths.

In the case \( \rho > 1 \), corresponding to a heavy upper fluid, the system exhibits linear instability at long wavelengths, but transitions into alternating stable/unstable zones as surface tension competes with gravity. A critical vertical asymptote appears at \( \rho \simeq 5.8275 \), beyond which capillary forces are insufficient to stabilize the dynamics at large wave numbers.

In the course of the study, it was also found that surface tension \( T \) plays a nontrivial and geometry-dependent role: for intermediate lower-layer depths, increasing \( T \) can either compress or expand regions of modulational stability. This behavior highlights the complexity of capillary effects and underlines the necessity for further numerical and analytical investigations of modulational phenomena in stratified fluid systems.
\\\\
{\bf Acknowledgement}
Dr. Olga Avramenko expresses gratitude to the Research Council of Lithuania for the support in preparing this article.

\end{document}